\documentclass[fleqn,usenatbib,useAMS]{mnras}

\usepackage{graphicx}	
\usepackage{amsmath}	
\usepackage{amssymb}	

\usepackage{xspace}
\usepackage{multirow}

\usepackage[dvipsnames]{xcolor}

\newcommand{\dd}{\mathrm{d}}

\newcommand{\pow}[1]{\ifmmode{}^{#1}\else ${}^{#1}$\fi}

\newcommand{\cm}{\,\ifmmode{{\mathrm{cm}}}\else cm\fi}

\newcommand{\ergps}{\,{\rm erg}\,{\rm s}\ifmmode{}^{-1}\else${}^{-1}$\fi}
\newcommand{\Mpch}{\,{\rm Mpc}\,\ifmmode h^{-1}\else $h^{-1}$\fi}
\newcommand{\snru}{\,\ifmmode{\mathrm{Myr}^{-1}}\else Myr${}^{-1}$\fi}
\newcommand{\kms}{\,\ifmmode{\mathrm{km}\,\mathrm{s}^{-1}}\else km\,s${}^{-1}$\fi\xspace}

\newcommand{\new}[1]{#1}

\newcommand{\HI}{{\text{H\MakeUppercase{\romannumeral 1}}}\xspace}

\newcommand{\Lya}{\ifmmode{\mathrm{Ly}\alpha}\else Ly$\alpha$\xspace\fi}
\newcommand{\coda}{\texttt{CoDaII}\xspace}
\newcommand{\msol}{{\rm M_{\odot}}}
\newcommand*\samethanks[1][\value{footnote}]{\footnotemark[#1]}

\usepackage[T1]{fontenc}
\usepackage{ae,aecompl}

\title[\Lya transmission in CoDaII]{Lyman-$\alpha$ transmission properties of the intergalactic medium in the CoDaII simulation}

\author[M. Gronke et al.]{Max Gronke${}^{1,2}$\thanks{E-mail: maxbg@jhu.edu}\thanks{Hubble fellow},
  Pierre Ocvirk${}^{3}$,
  Charlotte Mason${}^{4}$\href{Hfootnote.2}{\samethanks}, 
  Jorryt Matthee${}^{5}$,
  \newauthor
  Sarah E. I. Bosman${}^{6}$,
  Jenny G. Sorce${}^{7,8,9}$,
  Joseph Lewis${}^{3}$,
  Kyungjin Ahn${}^{10}$,
  \newauthor
  Dominique Aubert${}^{3}$,
  Taha Dawoodbhoy${}^{12}$,
  Ilian T. Iliev${}^{11}$,
  Paul R. Shapiro${}^{12}$,
  \newauthor
Gustavo Yepes${}^{13,14}$ 
\\
${}^{1}$Department of Physics, University of California, Santa Barbara, CA 93106, USA\\
${}^{2}$Department of Physics \& Astronomy, Johns Hopkins University, Baltimore, MD 21218, USA\\
${}^{3}$Observatoire Astronomique de Strasbourg, Université de Strasbourg, 11 rue de l’Université, 67000 Strasbourg, France\\
${}^{4}$Center for Astrophysics \,|\, Harvard \& Smithsonian, 60 Garden St, Cambridge, MA, 02138, USA\\
${}^{5}$Department of Physics, ETH Z\"urich, Wolfgang-Pauli-Strasse 27, 8093 Z\"urich, Switzerland\\
${}^{6}$Department of Physics and Astronomy, University College London, Gower Street, WC1E 6BT, London, UK\\
$^{7}$Univ Lyon, ENS de Lyon, Univ Lyon1, CNRS, Centre de Recherche Astrophysique de Lyon UMR5574, F-69007, Lyon, France\\
$^{8}$Univ Lyon, Univ Lyon1, ENS de Lyon, CNRS, Centre de Recherche Astrophysique de Lyon UMR5574, F-69230, Saint-Genis-Laval, France\\
$^{9}$Leibniz-Institut f\"{u}r Astrophysik (AIP), An der Sternwarte 16, D-14482 Potsdam, Germany\\
${}^{10}$Department of Earth Sciences, Chosun University, Gwangju 61452, Korea\\
${}^{11}$University of Sussex, Falmer, Brighton BN1 9QH, United Kingdom\\
${}^{12}$ Department of Astronomy, The University of Texas at Austin, TX 78712-1205, USA\\
$^{13}$Departamento de F\'{\i}sica Te\'orica M-8, Universidad Aut\'onoma de
Madrid, Cantoblanco, E-28049 Madrid, Spain.\\
$^{14}$Centro de Investigaci\'on Avanzada en F\'{\i}sica  Fundamental
(CIAFF), Universidad Aut\'onoma de Madrid, E-28049 Madrid, Spain.
}

\date{Draft from \today}

\pubyear{2020}

\begin{document}
\label{firstpage}
\pagerange{\pageref{firstpage}--\pageref{lastpage}}
\maketitle

\begin{abstract}
  The decline in abundance of Lyman-$\alpha$ (Ly$\alpha$) emitting galaxies at $z \gtrsim 6$ is a powerful and commonly used probe to constrain the progress of cosmic reionization.
  We use the \texttt{CoDaII} simulation, which is a radiation hydrodynamic simulation featuring a box of $\sim 94$ comoving Mpc side length, to compute the Ly$\alpha$ transmission properties of the intergalactic medium (IGM) at $z\sim 5.8$ to $7$. Our results mainly confirm previous studies, i.e., we find a declining Ly$\alpha$ transmission with redshift and a large sightline-to-sightline variation.
  However, motivated by the recent discovery of \textit{blue} Ly$\alpha$ peaks at high redshift, we also analyze the IGM transmission on the blue side, which shows a rapid decline at $z\gtrsim 6$ of the blue transmission.
This low transmission can be attributed not only to the presence 
of neutral regions but also to the residual neutral hydrogen within ionized 
regions, for which a density even as low as $n_{\rm HI}\sim 10^{-9}\,\mathrm{cm}^{-3}$ (sometimes combined with kinematic effects) leads to a significantly reduced visibility.
Still, we find that $\sim 1\%$ of sightlines towards $M_{\mathrm{1600AB}}\sim -21$ galaxies at $z\sim 7$ are transparent enough to allow a transmission of a blue Ly$\alpha$ peak.
  We discuss our results in the context of the interpretation of observations.
\end{abstract}


\section{Introduction}
\label{sec:intro}

While baryonic astrophysics in today's Universe is mainly governed by effects which are small-scale -- for cosmological standards -- this was not always the case.
A few billion years ago, at $z\sim 6$ the `Epoch of Reionization' (EoR) was nearly complete. During this period, the vast majority of atoms in the Universe underwent the same transition: from neutral to ionized \citep[for reviews, see e.g.][]{Robertson2010,2016ASSL..423.....M,2018PhR...780....1D}.
While this approximate picture stands on relatively firm grounds -- largely owing to the measurement of the Thompson optical depth by cosmic microwave background (CMB) experiments \citep{WMAP,Planck} -- the details are yet unclear. This is because the study of the EoR faces a number of challenges. On the observational side, because this period is so far away, thus, major obstacles do exist, for instance, for current and future $21$cm experiments.

But also on the theoretical side, the study of the EoR is fairly challenging. Not only is it, as mentioned above, a baryonic process and consequentially rather messy -- but it also spans a wide range of scales: the ionizing photons have to be produced (which can involve elaborate stellar modeling; e.g., \citealp{Eldridge2016}), then escape their birth cloud \citep[e.g.,][]{Kimm2019,Kakiichi2019}, and eventually the galaxy as well as its immediate surroundings \citep[e.g.,][]{1994ApJ...423..196D,Paardekooper2015,lewis2020}; each of these steps requires in principle knowledge of sub-parsec baryonic physics. Afterwards, the ionizing radiation might traverse $\gtrsim $Mpc sized already ionized regions in order to affect neutral atoms far away from the source. Thus, requiring the consideration of truly cosmological scales.

In recent years, however, progress has been made on both frontiers. The observational constraints of the EoR are improving continuously. Apart from the previously mentioned integrated CMB constraints, quasar \citep{2006AJ....132..117F,2015PASA...32...45B,2018MNRAS.479.1055B,2018ApJ...864..142D}, and gamma-ray bursts \citep{2013ApJ...774...26C} spectra yield tighter, and tighter constraints. Measurements of galactic emission help to constrain the evolution of the ionized fraction better -- mainly thanks to the Lyman-$\alpha$ (\Lya) line \citep[for a review, see][]{Dijkstra2014a}.
The change in the \Lya equivalent width distribution, and the change in the clustering statistics of \Lya emitters are nowadays commonly used as astrophysical EoR constraints \citep{2006MNRAS.365.1012F,2007MNRAS.381...75M,2016MNRAS.463.4019K,Mason2017,2018PASJ...70S..13O}.

This progress is built upon arduous observational work. While for decades, the detection of extra-galactic \Lya radiation was only conjectured \cite[][]{1967ApJ...148..377P,1980PASP...92..537K} later hundreds of \Lya emitting galaxies (LAEs) were detected \citep[e.g.,][]{1993A&A...270...43M,1998ApJ...502L..99H,2000ApJ...545L..85R}. Now this field is pushed towards higher redshift, and a continuously increasing amount and quality of \Lya data is available to the community
\citep[e.g,][]{2010ApJ...725L.205F,2018ApJ...864..103J,2018MNRAS.476.4725S,2019ApJ...878...12H}.

One particularly interesting development driven by new instruments and telescopes, is the availability of high-resolution, high-signal-to-noise \Lya spectra at $z\gtrsim 5$ -- unthinkable even just a decade ago. Thanks to this progress, it is now possible to study \Lya spectral properties even at these highest redshifts, which lead to the somewhat surprising detection of \textit{blue} \Lya peaks at $z\gtrsim 6$ \citep{Hu2016,Songaila2018,Matthee2018,Bosman2019}. Surprising, because prior to that it was commonly assumed (e.g., in models constraining the progress of the EoR) that all flux blueward of the \Lya line is absorbed at this high-$z$ due to the high neutral fraction of the IGM. Studying the detectability of blue peaks is one of the main focus of this work. 

Not just on the observational front but also computationally the EoR community is making steady progress due to the ever increasing power of supercomputers. Several groups manage now to run hydrodynamical cosmological volume boxes with radiative transfer to follow the evolution of the ionizing regions closely (e.g., on the fly in \citealp{CROC,2018MNRAS.479..994R} or in post processing in \citealp{2019MNRAS.485L..24K}).
This allows the community now to predict the evolution of observables as well as to estimate the scatter around these trends.
One recent example of such simulations is the \coda \citep{CoDaII} simulation which uses the \texttt{RAMSES-CUDATON} code \citep{CoDaI}.
 
In this work, we use the \coda simulation in order to study the evolution of \Lya observables at $6 \lesssim z \lesssim 7$.
This simulation features full coupled ionizing radiative transfer as well as a representative volume to study the \Lya observables at hand.
In particular, we focus on the disappearance of Lyman-$\alpha$ emitting galaxies (Lyman-$\alpha$ emitters; LAEs) at $z \gtrsim 6$ due to the increasing neutral IGM, and how this intergalactic absorption affects the \Lya line shape. In contrast to earlier studies, we focus in particular on the detectability of blue peaks at these redshifts -- motivated by the recent observations mentioned above.

This paper is structured as follows: in Sec.~\ref{sec:method} we lay out our method, that is, we provide technical background of the \coda simulation (\S~\ref{sec:coda}), and explain how we generate the transmission curves (\S~\ref{sec:transmission}). In Sec.~\ref{sec:result}, we present our results, which we discuss in some context in Sec.~\ref{sec:discussion}, before we conclude in Sec.~\ref{sec:conclusion}.

\begin{figure*}
  \centering
  \includegraphics[width=\linewidth]{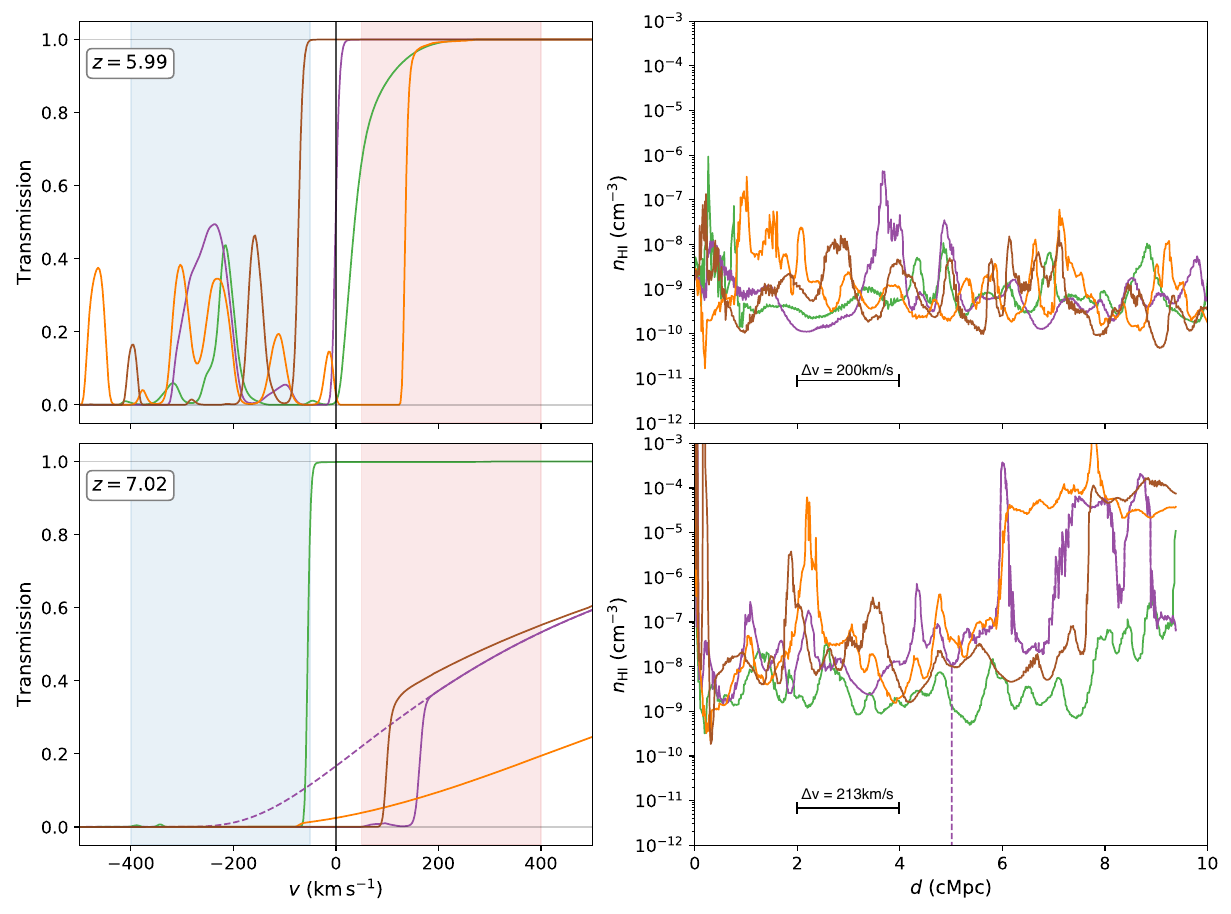}
  \caption{Examples of transmission curves (left panels) and the corresponding neutral hydrogen number densities (right panels) of an arbitrary halo at $z\sim 6$ (top row) and $z\sim 7$ (bottom row) with $R_{200}\sim 0.3\,$cMpc ($M\sim 2\times 10^{11}\,M_\odot / h$, $M_{\mathrm{1600AB}}\sim -22$ at $z\sim 6$). The purple dashed lines in the lower row show an example where we set $n_{\HI}(<\,5\,\mathrm{cMpc})=0$ for illustration purposes.
    The blue and red shaded region in the left column mark the $50 < |v|/(\kms) < 400$ region which we use as `blue' and `red' side in the further analysis.
    In the right panels, we mark the velocity corresponding to the Hubble flow of $\sim 2\,$cMpc at that redshift.
  }
  \label{fig:example_sightlines}

\end{figure*}

\begin{figure}
  \centering
  \includegraphics[width=\linewidth]{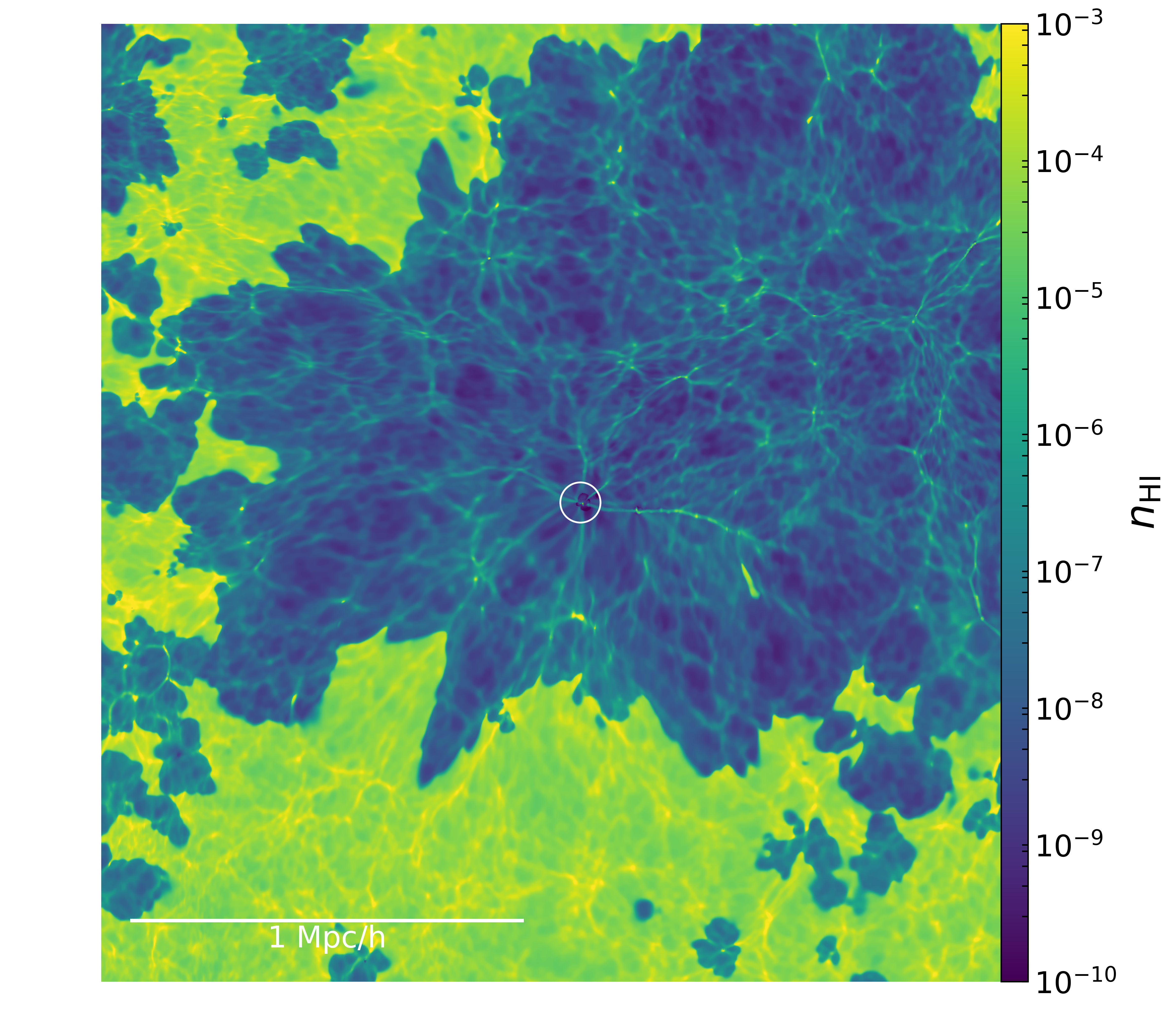}\\
  \includegraphics[width=\linewidth]{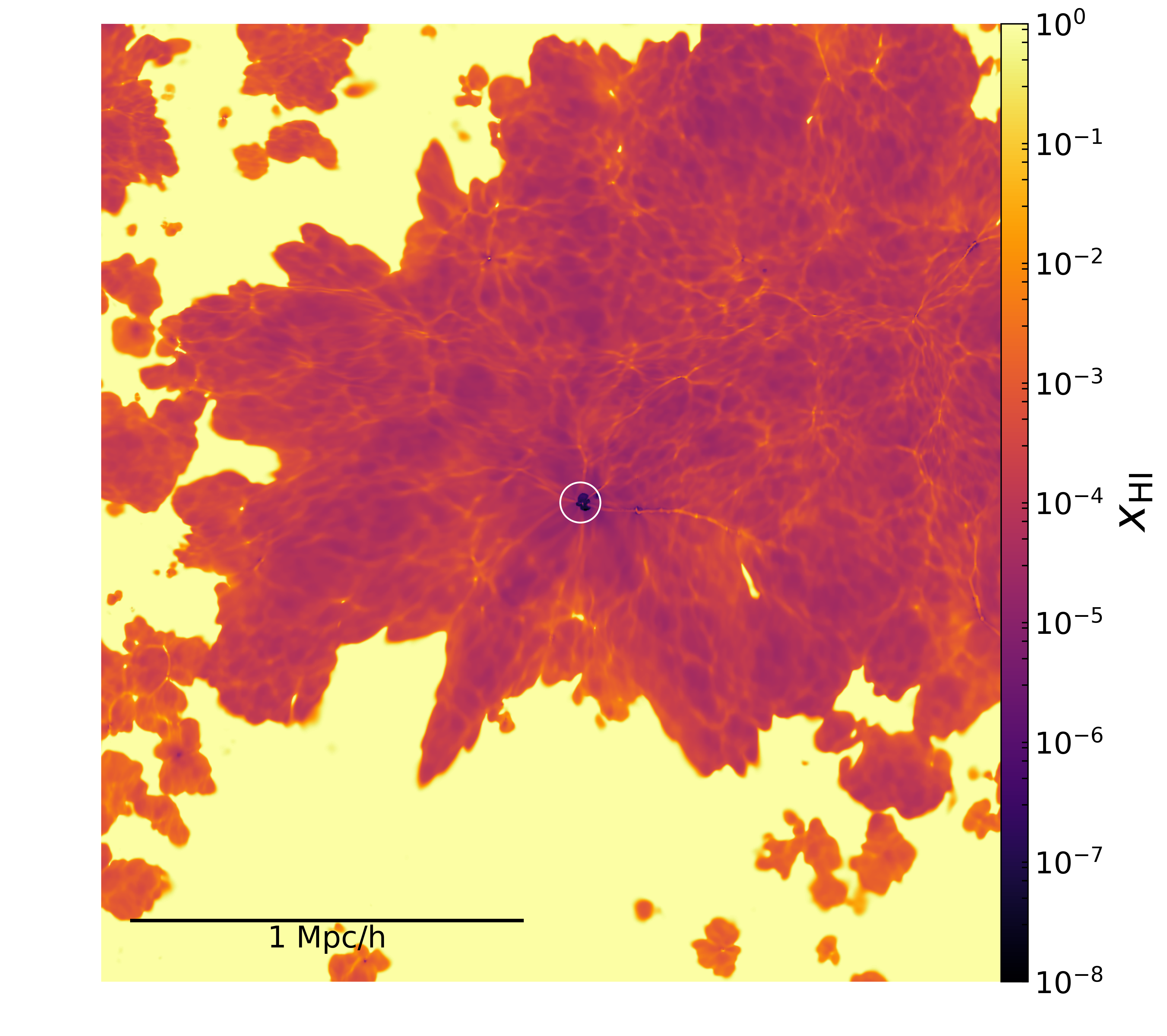}
  \caption{Slice through the simulation around one $M\sim 7\times 10^{11}\,M_\odot$ ($M_{1600AB}\sim -21.1$) halo at $z\sim 7$ (marked with a white circled indicating $2 R_{200}$). The upper and lower panel shows the neutral hydrogen number density and the neutral fraction, respectively. The plots illustrate the different ways of absorption discussed in \S~\ref{sec:example_sightlines}: \textit{(i)} wing absorption by the large, remaining neutral regions; \textit{(ii)} absorption through the neutral patches; or \textit{(iii)} resonant absorption by the residual neutral hydrogen in the otherwise ionized regions.}
  \label{fig:slice2d}
\end{figure}

\begin{figure}
  \centering
  \includegraphics[width=\linewidth]{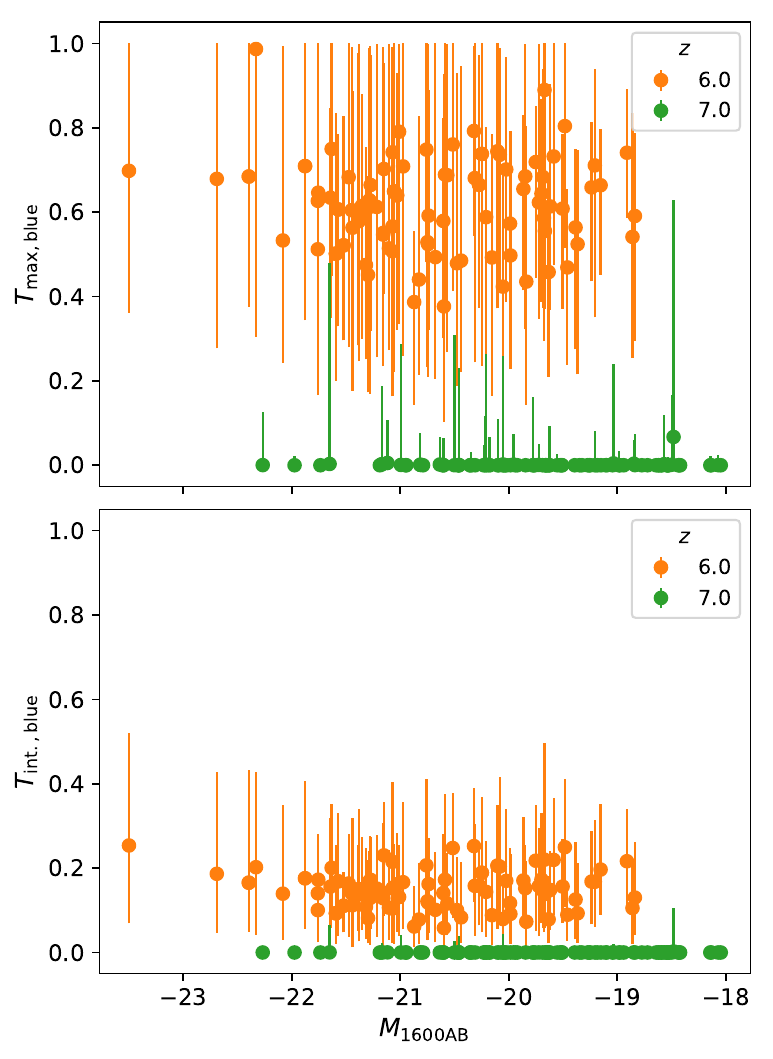}
  \caption{IGM transmission on the blue side of $100$ randomly selected halos. Each point and error bars represent the median, $16$th and $84$th percentile drawn from $100$ sightlines for each halo.
  }
  \label{fig:transmission_blue_overview}
\end{figure}

\begin{figure}
   \centering
  \includegraphics[width=\linewidth]{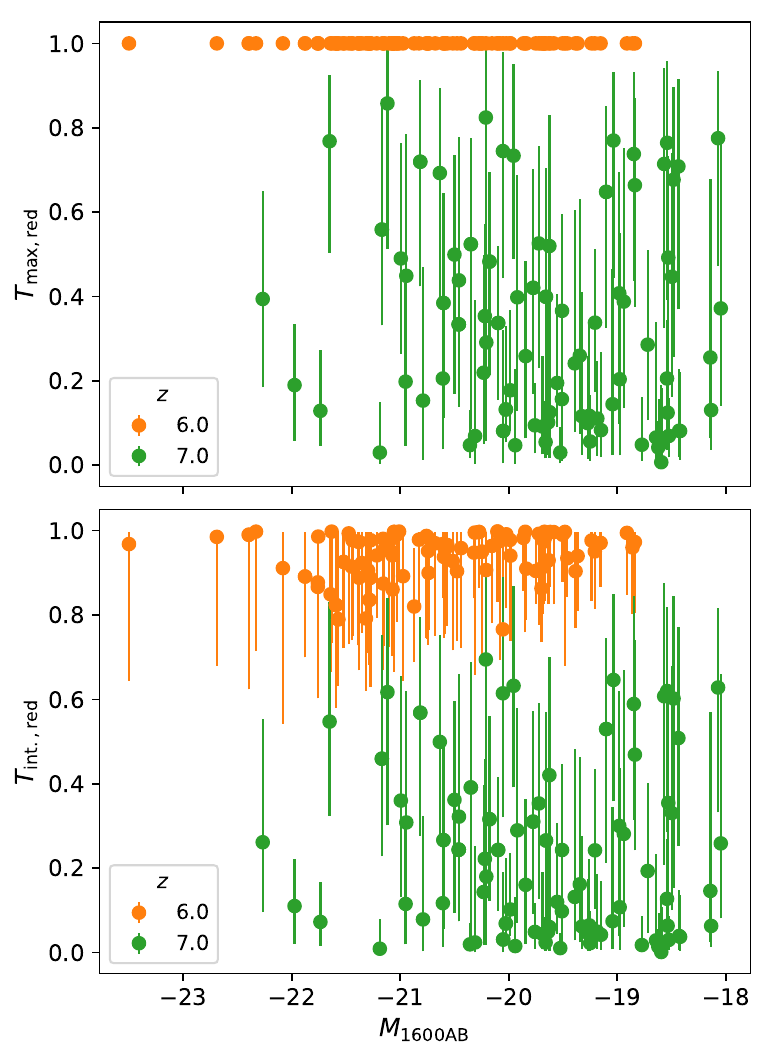}
  \caption{IGM transmission on the red side of the $100$ randomly selected halos. Each point and error bar
s represents the median, $16$th and $84$th percentile drawn from $100$ sightlines for each halo.}
  \label{fig:transmission_red_overview}
\end{figure}

\begin{figure*}
  \centering
  \begin{minipage}{.9\textwidth}
    \includegraphics[width=\linewidth]{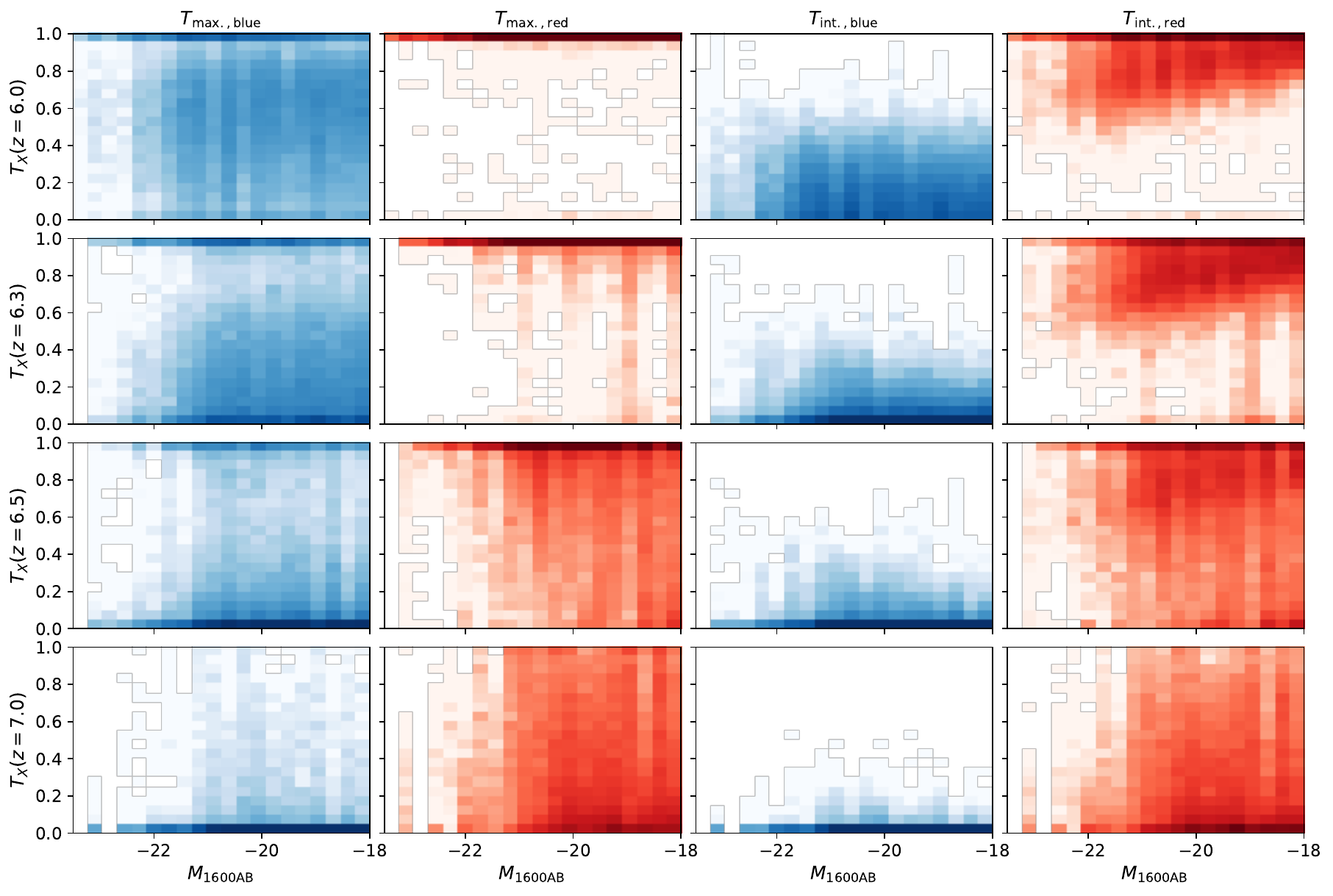}
  \end{minipage}
  \begin{minipage}{0.07\textwidth}
      \begin{tabular}{c}
        \includegraphics[width=.9\linewidth]{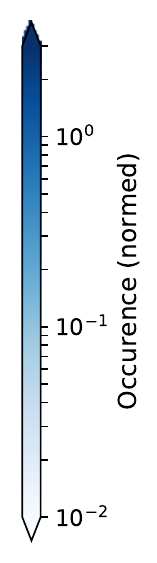} \\
        \includegraphics[width=.9\linewidth]{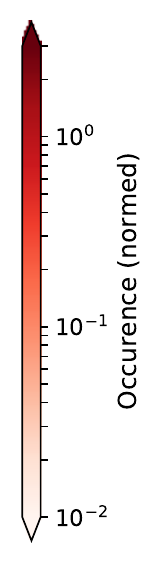}
      \end{tabular}
    \end{minipage}

  \caption{Distributions of the transmission properties. Shown are the maximum transmission on the blue and red sides (first and second column, respectively), and the normalized integrated transmission on both sides (third and forth column). Each row represents a redshift as denoted on the left side. For representation purposes, each 2D histogram is normalized and the color coding is logarithmically scaled.}
  \label{fig:transmission_2dhist}
\end{figure*}   
\section{Method}
\label{sec:method}

\subsection{The \coda simulation}
\label{sec:coda}
The \coda simulation is a fully coupled radiation-hydrodynamical simulation of galaxy formation during the Epoch of Reionization. It was performed on Titan, at Oak Ridge National Laboratory, with \texttt{RAMSES-CUDATON} \citep{CoDaI}. It is fully presented in \cite{CoDaII} and therefore we recall here only the main points relevant to this study. 
The simulation describes the evolution of a box of $\sim 94.5$ comoving Mpc (cMpc) on a side, i.e. large enough to model global reionization, from $z=150$ down to $z=5.8$ \citep[see, however,][for a discussion of the required boxsize in order to capture EoR fluctuations]{2014MNRAS.439..725I}. The simulation grid is $4096^3$, allowing us to resolve haloes down to $10^8 \msol$ (with a dark matter particle resolution of $\sim 4\times 10^5\msol$) and therefore providing a good sampling of the various halo masses contributing to cosmic reionization, as shown in \cite{lewis2020}. Another advantage of the \coda simulation is that it contains a statistically representative population of massive haloes, up to $10^{12} \msol$ at $z=6$, owing to its very large volume. Since the mass of dark matter haloes in which objects with blue \Lya peak mentioned above reside is unknown, it is crucial that our methodology allows us to investigate such a broad range of masses. 

Also, \coda reproduces a number of observables of the high redshift Universe, in particular its reionization history (i.e. the evolution of ${\rm x_{HII}}$ with redshift), and the UV luminosity functions at $z=6$ and above, which are particularly important for the investigations presented in this paper.
\new{Finally, the spatial resolution (better than $3.3\,$kpc physical) is well suited to computing \Lya transmission spectra. 
However, such a cell size may seem to allow resolving self-shielded systems only marginally. Indeed, \citet{2018MNRAS.478.5123R} showed that such systems are typically $1-10\,$pkpc at $z=6-7$. 
It is therefore possible that we are missing a contribution to opacity from those self-shielded systems below our resolution limit. While we can not offer a clear way of quantifying such potential missing opacity from the simulation, we 
can try to gain some insight into this aspect from the resolution study of \citet{2019A&A...626A..77O}, figure 7 and table 1. The latter quantifies the increase of residual neutral fraction after overlap in simulations with twice and four times the spatial resolution of \coda, i.e. with $8$ and $64$ times higher mass resolution. The most resolved simulation in that study has $0.92\,$pkpc cell size, i.e. marginally resolving the minimum size quoted by \citet{2018MNRAS.478.5123R}.  If self-shielded systems are very sensitive to resolution and hold a large amouint of the global \HI, we should see a significant increase in the residual $x_\HI$ when increasing resolution. Instead, the increase found is rather limited (about a factor of 2), despite increasing the mass resolution by a factor 64. Following this argument, the amount of \HI to account for unresolved self-shielded systems in \coda could also be an order unity effect. Generally, combining large volumes with spatial resolution high enough to capture self-shielded systems remains a challenge, and thus also a limitation of our methodology. Cosmic Dawn III will provide interesting insight into this aspect, with a factor of two increase in resolution and an improved calibration (Lewis et al., in prep.).}

\coda performs explicit radiative transfer of ionizing radiation from one cell to another, across the
simulation volume.  However, the amount of ionizing radiation released by each star particle
into the cell in which it forms was assumed to be reduced from the intrinsic photon luminosity of its stars
by the bound-free Lyman continuum opacity of the unresolved, subgrid-scale interstellar birth-cloud 
of the stars.  To account for this extra opacity, we adopted a fixed birth-cloud escape fraction of 
$f_{\mathrm{LyC}}=0.42$ which the intrinsic stellar luminosity of each star was reduced when assigning an
ionizing photon luminosity to the stars in each star particle.
This value was tuned to reproduce a range of observables of the global ionization history, although it somewhat underpredicts the neutral fraction at the tail end of the EoR \citep[see][for a full discussion]{CoDaII}.
Notably, \coda does use the full speed of light in its ionizing radiative transfer routines, and, thus, does not suffer from problems stemming from the usage of the `reduced speed of light approximation' \citep[in combination with the M1 closure relation][]{2019A&A...626A..77O,2019A&A...622A.142D}.

For the generation of the UV magnitudes cited ($M_{\mathrm{1600AB}}$) the BPASS $Z = 10^{-3}$ binary population and spectral synthesis model was used \citep{2017PASA...34...58E}, assuming no dust extinction.

\subsection{Generation of transmission curves}
\label{sec:transmission}

We analyzed the simulation snapshots at $z\sim\{5.8,\,6,\,6.26,\,6.55,7\}$ (with average neutral fractions of $\log_{10}(\langle x_{\HI} \rangle)\sim \{-5.15,\,-4.92,\,-1.41,\,-0.64,\,-0.30\}$). At each of these snapshots, we randomly selected $50$ halos for each $0.5$ magnitude bin with $M_{\mathrm{UV}}< -18$. For each halo, we drew random lines of sight for which we generated transmission curves around the \Lya wavelength. In particular, we
\begin{enumerate}
\item cut out a spherical region $< 1.5 R_{200}$ (i.e., set all the cells within that radius to be fully ionized) around the halo position (as given by the halo finder) where $R_{200}$ is the
  radius in which the average density is $200$ times the mean matter density at that redshift \citep[as defined in ][equation~2]{CoDaII}. We chose the cutout of $1.5 R_{200}$ for a number of (connected) reasons. The main reason is that the circumgalactic medium (CGM) is believed to span $1-2$ virial radii around each galaxy \citep[e.g.,][]{Tumlinson2017} and current cosmological simulations are not able to resolve the cold, neutral gas within it \citep{Fielding2016,Liang2016}, in fact, it has been shown that the \HI content of such simulations in the CGM is non-converged \citep{VandeVoort2018,Hummels2018,Peeples2018,Suresh2018}. Since \Lya radiative transfer is dependent on small-scale structure within the \HI \citep[e.g.,][]{Neufeld1991,Gronke2017} this non-convergence as well as underresolving these structures is highly problematic when performing a full \Lya radiative transfer. We, thus, treat the radiative transfer processes within the interstellar and circumgalactic medium as a `black box', and concentrate on the impact of the IGM. The second reason for the cutout is a more technical reason. Even if we could resolve all the \HI structure within in the ISM \& CGM perfectly, photons scattered within $\lesssim 1.5 R_{200}$ have a non-negligible probability to scatter back into the line-of-sight \citep{Laursen2011,Jensen2014}, i.e., we could not treat the scattering process as absorption.
\item we compute the gas mass weighted mean velocity within $< R_{200}$ which we use as systemic redshift for each halo, that is, effectively shifting all velocities with respect to this one. 
\item we drew $100$ random sightlines around each halo \new{not intercepting any domain boundary}. Along each, we calculate the transmission curve using \texttt{trident} \citep{Hummels2017} taking the neutral hydrogen number density $n_\HI$, the temperature $T$, and the gas velocity $\mathbf{v}$ for each cell into account.
We integrate from the halo position to a distance of \new{$5000\kms/H(z)$} away from the source which is sufficient to have converged transmission curve \new{\citep[see, e.g., figure 1 of][]{2020arXiv200413065M}}. We tested this assumption by finding that a subset of the transmission curves were unchanged if we lowered this threshold to \new{$\sim 2000\kms$}.
\end{enumerate}
In summary, this procedure allows us to be agnostic about the intrinsic line shape emergent from the \Lya emitting galaxy.

\begin{figure}
  \centering
  \includegraphics[width=\linewidth]{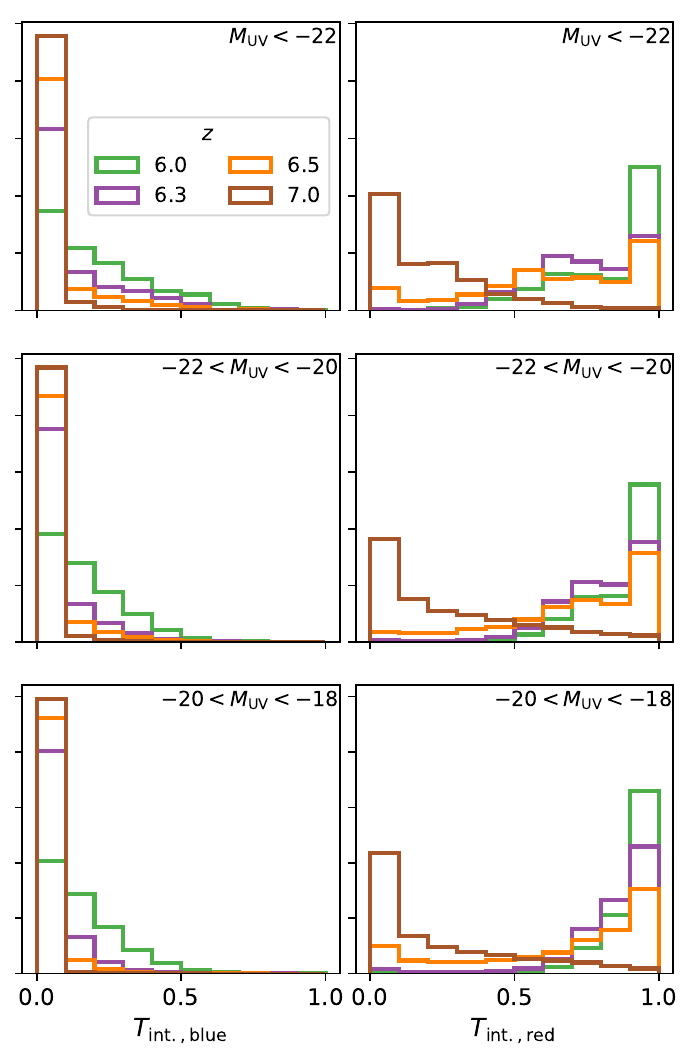}
  \caption{Normalized sightline distribution of integrated transmission split by UV magnitude and red / blue side. The color coding corresponds to different redshifts.}
  \label{fig:transmission_pdf_multiplot}
\end{figure}

\begin{figure}
  \centering
  \includegraphics[width=\linewidth]{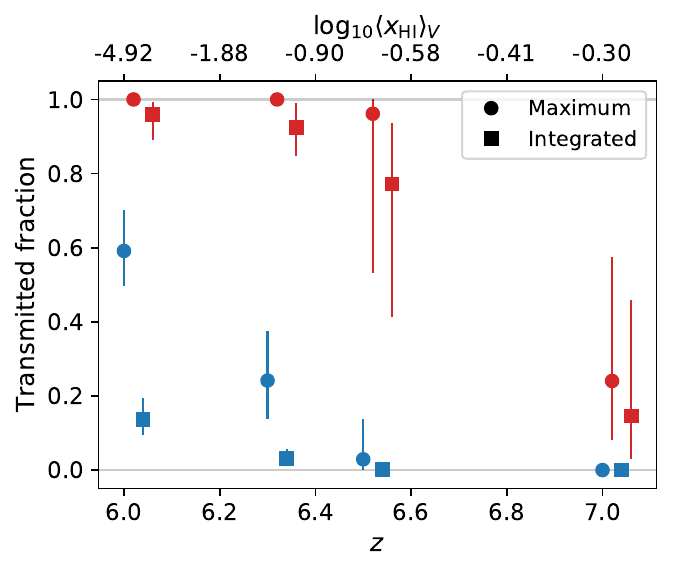}
  \caption{Evolution of transmitted maximum and integrated flux (circles and squares, respectively) on the red and blue side (in the corresponding color) \new{of all the selected halos}. Each point and error bars show the $16$th, $50$th and $84$th percentile of the medians, i.e., represent the scatter between the sightlines. The secondary $x$-axis on top of the plot shows the volume averaged neutral fraction of the \coda simulation at the respective redshifts.
  }
  \label{fig:transmission_evolution}
\end{figure}

\begin{figure}
  \centering
  \includegraphics[width=\linewidth]{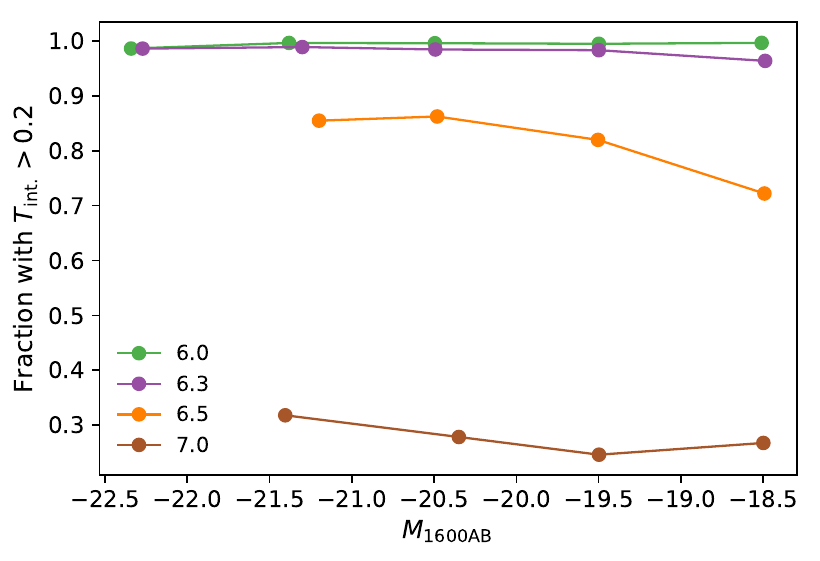}
  \caption{Evolution of fraction of halos with total $T_{\rm int.} > 0.2$ (i.e., $|v|<400\kms$). Shown are only the $M_{\mathrm{UV}}$ bins with at least $10$ halos.}
  \label{fig:lae_fraction}
\end{figure}

\section{Result}
\label{sec:result}

\subsection{Illustrative example sightlines}
\label{sec:example_sightlines}
Figure~\ref{fig:example_sightlines} shows some arbitrary transmission curves originating from a randomly chosen halo (left column) and in  corresponding color in the right column the neutral hydrogen number density as a function of distance from the source. This illustrates how \Lya photons are absorbed in the IGM. At $z\sim 7$ (lower row in Fig.~\ref{fig:example_sightlines}), for instance, one notices the characteristic wing of the absorbing Voigt profile at $v \gtrsim 200\kms$. Furthermore, though, the transmission curves clearly exhibit some resonant absorption closer to $v\sim 0$. This is in spite of the fact that the edge of the ionized bubble is in this example for most sightlines at $d \sim 6 \mathrm{cMpc}\sim 640\kms / H(z)$, i.e., too far shifted to be responsible for the resonant absorption. However, as can be seen in the right panel, inside this ionized region, patches of $n_{\HI}\gtrsim 10^{-8}\cm^{-3}$ exist which are responsible for the resonant absorption. These patches can be fairly close to the emitting galaxy, and thus, likely infalling causing absorption even on the  red side of the spectrum \citep[][]{Dijkstra2007a,2008MNRAS.391...63I}.

These patches with a ``large'' \HI number density of $n_\HI \gtrsim 10^{-8}\cm^{-3}$ can cause resonant absorption also at lower redshifts which can be seen in the upper panels of Fig.~\ref{fig:example_sightlines}. In these examples, the patches lead to a rather noisy \Lya transmission curve on the blue side.

Apart for the neutral region, and neutral patches inside the ionized region, also the residual neutral fraction inside the ionized bubbles can cause significant absorption. We illustrate this in the lower row of Fig.~\ref{fig:example_sightlines} where we modified one sightline by setting $n_\HI(< 5\,\mathrm{cMpc})=0$. The solid and dashed purple curve in Fig.~\ref{fig:example_sightlines} shows the unmodified and altered sightline, respectively. Clearly the region of wing absorption ($\gtrsim 200 \kms$) is unchanged. However, due to the residual neutral part inside this region the spectrum blueward of $\sim 200\kms$ shows zero transmission. 

That even a small neutral hydrogen number density is sufficient in order to cause significant resonant absorption can be seen by this simple estimate:
\begin{equation}
  \label{eq:fiducial_neutral_estimate}
  n_{\HI}\sim \frac{\tau}{d \sigma_0} \sim \tau \frac{H(z)}{\sigma_0 v_{\mathrm{core}}}
\end{equation}
where $\sigma_0$ and $v_{\mathrm{core}}$ are the \Lya cross section at line center and the width of the core region which take values of $\sim 5\times 10^{-14}\cm^{2}$ and $\sim 80\kms$ at $T\sim 10^4\,\mathrm{K}$, respectively \citep[e.g.,][]{Dijkstra2017}. For $z\sim 6$ ($z\sim 7$) this yields $n_{\HI}\sim 2\times 10^{-10}\tau\cm^{-3}$ ($\sim 4\times 10^{-10} \tau\cm^{-3}$). Equation~(3) in \citet{1997ApJ...490..564W} is a more precise estimate which yields similar values. In Appendix~\ref{sec:fid_neutral_frac} we investigate this numerically.

Figure~\ref{fig:slice2d} illustrates the three different absorption causes. It shows a slice around a halo at $z\sim 7$. Clearly, the large remaining neutral regions are visible. Furthermore, also the remaining neutral absorbers inside the ionized regions can be seen. Most importantly, though, the neutral hydrogen number density even inside the ionized regions shows a lot of structure with only small patches exhibiting $n_{\mathrm{HI}}\lesssim 10^{-9}\cm^{-3}$ required for a non-absorption at this redshift.

\subsection{\Lya transmission statistics}
\label{sec:transmission_statistics}
Figure~\ref{fig:transmission_blue_overview} shows the median and difference to the $16$th and $84$th percentile of the maximum transmission $T_{\mathrm{max,blue}}$ within $v\in [-400,\,-50]\kms$ which we define as the ``blue side''. 
The lower panel of Fig.~\ref{fig:transmission_blue_overview} shows the integrated transmission in this wavelength range $T_{\rm int. blue}\equiv \int T(v)\dd v / \Delta v_{\rm blue}$. Several points are clear from Fig.~\ref{fig:transmission_blue_overview}: \textit{(i)} at $z \gtrsim 7$ essentially all \Lya transmission emergent on the blue side of the spectrum is absorbed by the IGM in the \coda simulation with some rare peaks being transmitted, \textit{(ii)} at $z\sim 6$ some transmission on the blue side is allowed, \textit{(iii)} but even at this later times there is a strong sightline-to-sightline variation for all the halos, and \textit{(iv)} overall, there seems to be no clear dependence on the UV magnitude.

Fig.~\ref{fig:transmission_red_overview} shows the same statistics but for the red side ($v\in [50,\,400]\kms$) of the \Lya transmission curve. As expected, the overall transmission is much larger with a $T_{\mathrm{max,red}}\sim 1$ for $z \lesssim 6$, and even the integrated transmission reaching mostly $\gtrsim 80\%$ of its maximum value at $z\sim 6$. Also at higher redshifts a large fraction of the redward flux is transmitted reaching $\sim 0-80\%$ at $z\sim 7$.
As for the blue side, we found a rather large sightline-to-sightline variation.

Note that in both Fig.~\ref{fig:transmission_blue_overview} and Fig.~\ref{fig:transmission_red_overview} we show only $100$ randomly selected halos for visualization purposes. Fig.~\ref{fig:transmission_2dhist} shows instead the full distribution for all the analyzed halos of these transmission statistics. Apart from the confirmation of the main findings stated above, the evolution in the transmission properties is clearly visible.
In addition, there seems to be a slight brightness dependence of $T_{\mathrm{int., red}}$ with the more massive halos having a larger probability that their \Lya is not transmitted on the red-side. We attribute this effect to the larger cosmological infall velocity (with respect to the (\Lya) emitting regions) for these halos which leads to an increased absorption on the red side (see \S~\ref{sec:example_sightlines}).
Though, note that Lya from these massive halos may be emitted anyway with a larger velocity offset due to more scattering in the ISM \citep{2014ApJ...785...64S,2017MNRAS.464..469S,Mason2017}.

Another representation of the integrated transmission is shown in Fig.~\ref{fig:transmission_pdf_multiplot} -- which, assuming an intrinsic \Lya equivalent width (EW) distribution and respective \Lya escape fraction -- could be translated to an observed EW distribution of Lyman break galaxies. On the red side, the evolution from a bimodal distribution at $z\sim 7$ to a unimodal distribution with most sightlines yielding $T_{\mathrm{int., red}}\sim 1$ at $z\sim 6$ can be observed. The integrated blue transmission shows a non-negligible tail with $T_{\mathrm{int.,blue}}>0$ which becomes more prominent at later times. As we discuss in \S~\ref{sec:observations} this tail is important for observed blue peaks at high-$z$.

Focusing on the red side (right column) of Fig.~\ref{fig:transmission_pdf_multiplot}, it is interesting that at earlier times ($z\sim 6.5$), the transmission statistics are more bimodal \citep[see also][]{2015MNRAS.446..566M}. I.e., the transmission is either very large or close to zero. This makes sense since at such high-$z$ the $T(v)$ is essentially a step function due to the large \HI cross section, and the fact that $T\propto e^{-\tau}$. Specifically, one can write $T(v)\sim 0$ for $v \lesssim v_{\mathrm{cutoff}}$ where $v_{\mathtt{cutoff}}$ is commonly set by the infall velocity; as discussed in \S~\ref{sec:example_sightlines}. This distribution might be relevant for searches of \Lya emitting galaxies which optimally should target as many sources as possible assuming a unity IGM transmission. At $z\gtrsim 7$, on the other hand, we find the integrated transmission peaks at $0$.

Figure~\ref{fig:transmission_evolution} summarizes the redshift evolution of these statistics. Here, we show the percentiles of the medians for every halo as a function of redshift. We show both the maximum and normalized integrated transmission. Fig. \ref{fig:transmission_evolution} shows that the decline on the blue side is rather rapid with no flux transmitted at $\gtrsim 6.5$. The decline on the red side, on the other hand, is offset by $\Delta z\sim 0.6$ towards higher redshifts.

In Fig.~\ref{fig:lae_fraction}, we show the evolution of the fraction of sightlines with a \textit{total} integrated (total) transmission $> 0.2$ (within $50<|v|/ (\kms) < 400$).
Since this observed equivalent width is a product of the intrinsic equivalent width, the galactic \Lya escape fraction, and the integrated intergalactic transmission, this can be understood as the impact of the IGM on the `\Lya emitter fraction'\footnote{The `\Lya emitter fraction' is defined as the fraction of Lyman break galaxies with an \Lya EW above some threshold -- usually $20\,$\AA. Our choice of $0.2$ as a cutoff is motivated by this, and a `common' intrinsic \Lya EW of $\sim 100\,$\AA \citep{2003A&A...397..527S}.}. 
Between $z\sim 6.5$ and $z\sim 7$ we see a rapid decline for all halo sizes. The most UV luminous halos show a slightly larger integrated transmission which is likely due to their larger virial radius, i.e., whether or not this signature holds will heavily depend on the CGM evolution (see the discussion \S~\ref{sec:caveats} on that matter) -- but could also be due to them residing in more ionized regions \citep{2018ApJ...857L..11M}.

\begin{figure*}
  \centering
  \includegraphics[width=.9\textwidth]{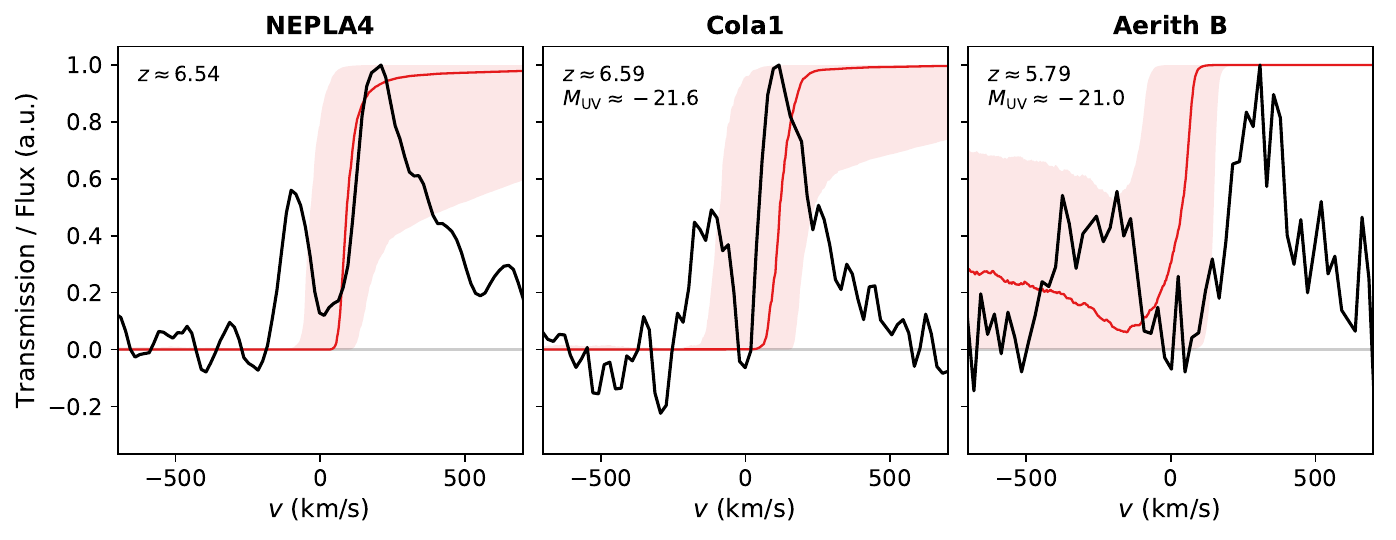}
  \caption{\Lya spectra of `NEPLA4', `COLA1', and `Aerith B' (from left to right panel) with the median (and 16th to 84th percentiles as shaded region) transmission curves of similar halos in \coda. Specifically, the central [right] panel shows sightlines (100 per halo, maximum of 200 halos) originating from halos within $\pm 0.25$ magnitude of the observed value at $z=6.55$ [$z=5.8$].}
  \label{fig:data_comparison}
\end{figure*} 

\begin{figure}
  \centering
  \includegraphics[width=\linewidth]{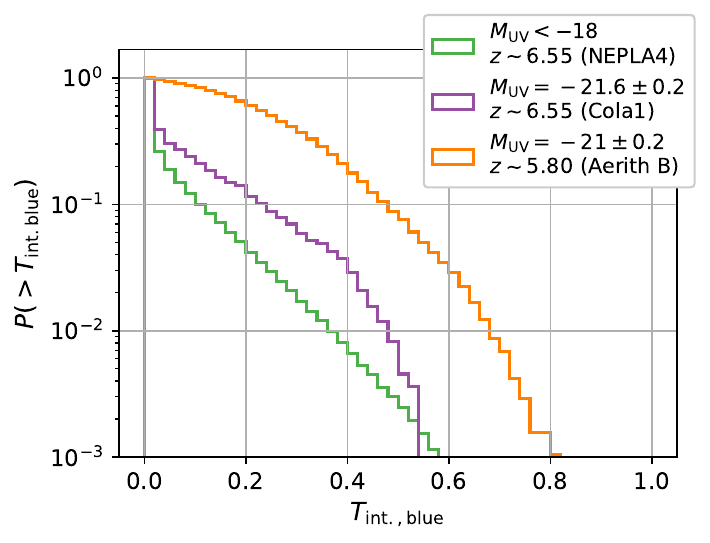}
  \caption{Inverse cumulative distribution function of the integrated flux on the blue side for halos similar to `COLA1', `Aerith B', and `EPLA4'.}
  \label{fig:cdf_data}
\end{figure}

\begin{figure}
  \centering
  \includegraphics[width=\linewidth]{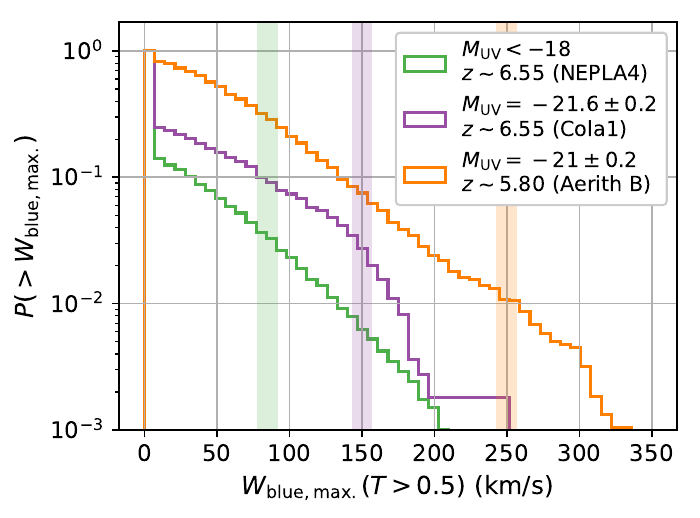}
  \caption{Inverse cumulative distribution function of the maximum width of a transmitted peak on the blue side.}
    \label{fig:cdf_data_width}
\end{figure}

\subsection{Comparison to observed blue peaks at high-$z$}
\label{sec:observations}

Recently, several \Lya spectra at $z \gtrsim 6$ with a prominent blue peak have been observed \citep{Songaila2018}. Specifically, `Aerith B' \citep{Bosman2019}, `COLA1' \citep{Hu2016,Matthee2018}, and `NEPLA4' \citep{Songaila2018} are well-studied examples which we want to compare to our findings.

`COLA1' is a luminous \Lya emitter at $z=6.59$ located in the well-studied COSMOS field. Besides being exceptionally bright in \Lya (L$_{\rm Ly\alpha} = 4\times10^{43}$ erg s$^{-1}$) it appears reasonably bright in the UV continuum with M$_{\rm UV}=-21.6\pm0.3$. Contrarily to other galaxies with similar UV luminosity, COLA1 appears particularly compact \citep{Matthee2018}.

`NEPLA4' is a \Lya emitter at $z\sim 6.54$ with a line shape resembling that of `COLA1'. Since its location is in the currently less-studied NEP field the rest-frame UV magnitude is, unfortunately, unknown.

Lastly, `Aerith B' is a bright ($M_{\rm AB} = -21.0_{-0.2}^{+0.3}$) Lyman-break galaxy at $z\approx 5.79$ displaying strong \Lya emission ($EW_{\rm rest} = 64^{+39}_{-24}$\AA). Aerith B is situated within the ionisation cone of a neighbourhing bright quasar at a distance $d\sim 750$ proper kpc, with an estimated resulting UV intensity at its location of $J_{21} = 406 \pm 40$: a factor $\sim 100$ higher than the cosmic peak of the UVB. 
Unlike COLA1 and NEPLA4, the velocity separation between the red and blue peak of the \Lya line is large in Aerith B ($\Delta v = (580\pm80)\kms$), indicating the galaxy is unlikely a significant Lyman continuum leaker \citep{2016ApJ...828...71D,2016MNRAS.461.3683I}.

Figure~\ref{fig:data_comparison} shows the spectra of these three objects \citep[specifically from][from left to right]{Songaila2018,Matthee2018,Bosman2019} alongside with transmission curves of `similar' halos, i.e., within $\pm 0.25$ UV magnitude of the $M_{\rm UV} \approx -21.6$ and
$M_{\rm UV} \approx -21.0$  `COLA1' and `Aerith B', respectively, have been associated with. In Fig.~\ref{fig:data_comparison} we show the median transmission curve as well as the $16$th and $84$th percentile of the distribution stemming of $100$ sightlines for the $25$ [$184$] halos falling in the right $M_{\rm UV}$ range for `COLA1' [`Aerith B']. It is clear that finding an object such as `COLA1' or `Nepla4' in the \coda simulation is extremely unlikely. On the other hand, while a \Lya spectrum such as shown by `Aerith B' is still far from common, it seems at least not entirely unlikely from Fig.~\ref{fig:data_comparison}. However, the median transmission curve can be misleading, and one should directly compare the impact of the IGM on the \Lya spectral properties (for a detailed discussion of this effect, see \citealp{Byrohl2020}).

In Fig.~\ref{fig:cdf_data} and Fig.~\ref{fig:cdf_data_width} we show the distribution of all the integrated flux on the blue side $F_{\mathrm{int., blue}}$ and the maximum width of a transmitted blue peak $W_{\mathrm{blue,max.}}(T>0.5)$ (defined as an uninterrupted transmission $T>0.5$), respectively. Specifically, we show the inverted cumulative distribution functions of the sightlines originating from similar halos as `COLA1', `Aerith B', and `NEPLA4'. Since, as described above, for the latter the UV magnitude is unknown, we use our full distribution at that redshift.

Figs.~\ref{fig:cdf_data} and \ref{fig:cdf_data_width} quantify the statement of ``extremely unlikely'' made above better. Given the blue peak height observed and making the -- quite optimistic -- assumption of an intrinsically symmetric spectrum, one requires $T\gtrsim 0.5$ on the blue so in order to explain the observations. Fig.~\ref{fig:cdf_data} shows that for such halos this occurs in $\sim 1-10\%$ of all the sightlines. Maybe more revealing is Fig.~\ref{fig:cdf_data_width} where we show the distribution of the maximum blue peak width transmitted at least $50\%$ by the IGM. In this figure, we also show as vertical lines the approximate blue peak width of the three observed objects studied. Due to the wide peak of `Aerith B' the probability of finding an `Aerith B' like object in \coda is $\sim 1\%$. Inversely, since `NEPLA4' shows a blue peak width of only $\sim 80\kms$, the probability of transmitting such a blue peak at this redshift is $\sim 10\%$. Note, again, however, that due to the lack of UV information we show the full
$W_{\mathrm{blue,max.}}(T>0.5)$ distribution at $z\sim 6$.
We also caution that firstly the assumption made of the intrinsic spectrum might be unlikely itself since most \Lya emitters at lower redshift possess a mostly \Lya spectrum asymmetric towards the red side \citep[e.g.,][]{Yang2015,Erb2014}, \new{and secondly the systemic redshift of the sources is unknown yielding the possibility of two red peaks being detected}. We discuss these caveats and the likelihood of observing more blue peaks (\new{or two red peaks}) at high-$z$ in \S~\ref{sec:disc_blueobs}.

Unlike COLA1 and NEPLA4, Aerith B is known a priori to be located in a very biased environment -- an ionised quasar proximity zone -- which is not captured by \coda due to the non-inclusion of quasars as ionizing sources. It is then interesting that the simulation predicts LAEs similar to Aerith B have a non-zero probability of being detected `in the field' at $z=5.8$. This may be related to the fact that CoDa II could overpredict J21 after overlap by a factor $\sim 10$, and similarly underpredicts the neutral fraction by the same factor \citep[see][figure 3]{CoDaII}. Despite observations of LAEs being more extensive at $z\sim 6$ than at $z\sim 6.5$, no other such objects are currently known.

\section{Discussion}
\label{sec:discussion}

\subsection{Implications of core-scatterings in an ionized IGM}
\label{sec:core_scatterings_implications}

As we showed in Sec.~\ref{sec:result}, absorption in the core of the line due to neutral hydrogen atoms in the ionized regions of the IGM is responsible for a large fraction of the absorbed flux. They lead to essentially zero transmission on the blue side, and -- due to cosmological infall -- can also heavily affect the red side. This means that there is no simple relation anymore between the size of the ionized region, and the absorption pattern, and has, thus, several implications for the study of the EoR, which we want to discuss here.

The simple picture of the size of the ionized bubbles, and the cutoff of the \Lya line linked to the edge of the bubble often put forward in the literature is not valid anymore if there is a significant ($n_{\HI}\gtrsim 10^{-10}\cm^{-3}$) neutral hydrogen component present. This is the case in the \coda simulation at $z \gtrsim 6.5$\footnote{The abundance of sinks of ionizing photons such as Lyman limit systems and the boxsize do affect these low neutral fractions \citep{2014MNRAS.439..725I}.}. 
Instead, the visibility is mainly set by the (size of the) highly ionized regions around \Lya emitting sources with $n_{\HI}\lesssim 10^{-10}\cm^{-3}$, and the kinematics of the gas surrounding these galaxies. Due to the lack of quasars or other high-energy sources in \coda, the presence of these regions is mainly set by the redshift. This can be seen by the rapid development of the transmission on the blue side (cf. \S~\ref{sec:transmission_statistics}).

This affects all the \Lya observables commonly used to constrain the EoR, that is, the \Lya equivalent width distribution, LAE fraction, \Lya luminosity function, and the clustering statistics of LAEs.

\subsection{Caveat: the omission of galactic and circumgalactic radiative transfer}
\label{sec:caveats}

The main caveat of this work, is the focus on the transmission of the intergalactic medium, i.e., leaving out the \Lya radiative transfer processes in the ISM and CGM. Evolution of galactic and circumgalactic properties, can change the \Lya escape fraction, and thus, lower the observed \Lya luminosity. In particular, a lower \Lya escape fraction might mimic a lower IGM transmission, that is, a higher IGM neutral fraction / redshift.

Since \Lya photons are primarily destroyed by dust,
two components can theoretically lower the \Lya escape fraction: \textit{(a)} a larger dust content, and \textit{(b)} a longer path length of \Lya photons (through this dusty medium). While an on average larger dust content towards higher redshift seems unlikely, the latter option might be feasible. For instance, a lower ionization background leads to a larger HI column density in the CGM, thus, increasing the path length of \Lya photons. This scenario was discussed in \citet{Sadoun2016} who manage to reproduce the observed drop in the \Lya emitter fraction \citep[e.g.,][]{Stark2011,2013ApJ...775L..29T,Schenker2014}.

However, while this scenario lowers the \Lya escape fraction, and consequently, the observed equivalent widths in agreement with observations, it also changes the other \Lya observables. In particular, a larger optical depth / path length leads to an increased frequency diffusion, and hence, wider \Lya lines \citep[e.g.,][]{Neufeld1990}. This seems in tension with observed \Lya spectra at high-$z$ \citep{2010ApJ...725..394H,2010ApJ...723..869O,2018A&A...619A.147P,2020MNRAS.492.1778M}
but further study is required to come to a firm conclusion.

While one solution to the caveat might seem the inclusion of the ISM \& CGM into the radiative transfer calculations, we deliberately chose not to do so in this study. As preluded in \S~\ref{sec:transmission}, this is mainly due to three reasons: \textit{(i)} \Lya radiative transfer is sensitive to sub-parsec structure inside the neutral hydrogen such as its clumpiness \citep[e.g.,][]{Neufeld1991,Gronke2017}, \textit{(ii)} this structure exists \citep[e.g., in the CGM][]{Rauch1999,Lan2017} and \textit{(iii)} \coda (as similar simulations) cannot resolve these scales in the ISM / CGM. In summary, while the ISM / CGM do have an effect on the \Lya line we do not resolve the relevant scales here, so we only consider the IGM effects.
In fact, the inability to reproduce \Lya spectra of radiative transfer simulations using high-resolution galactic simulations might be due to this  issue \citep[see discussion in][]{Gronke2017}.
Instead, we chose to follow a `Russian doll' approach to tackle the multiscale problem of modeling \Lya observables, and focus on the intergalactic part in this study.

\subsection{\textit{Allez les bleus:} The curious cases of blue \Lya peaks at high-$z$}
\label{sec:disc_blueobs}

In Sec.~\ref{sec:observations} we compared our findings to observations of blue \Lya peaks at $z\gtrsim 6$. Specifically, we analyzed the observed spectra of `Aerith B' \citep{Bosman2019}, `COLA1' \citep{Hu2016,Matthee2018}, and `NEPLA4' \citep{Songaila2018} (cf. Fig.~\ref{fig:data_comparison}). Other authors did claim a detection of a blue \Lya peak at high redshift such as the $\sim 450\kms$ blueshifted peak at $z\sim 9.1$ described in \citet{Hashimoto2018}. However, due to a lack of a (dominant) red peak, this is less certainly \Lya compared to the three objects we focused on here. Similarly, \citet{Songaila2018} present more `complex' \Lya profiles with a blue wing -- but not necessarily a clear blue peak. Better spectral resolution and sensitivity will help to clearly show which of these are blended peaks.

In the analysis presented in \S~\ref{sec:observations}, we concluded that an IGM transmission required in order to observe these cases occurs in only a few percent of the sightlines in \coda (mostly where motions allow for a blueward transmission).
The crucial assumption leading to this number is the intrinsic symmetry of the \Lya spectrum. For  lower redshift \Lya emitting galaxies, we know that in fact most \Lya lines are asymmetric towards the red side.
At $z\sim 0.3$ high-resolution spectroscopy of high-redshift analogs suggests that less than half of the flux is emerging on the blue side\footnote{For instance, the $12$ `Green peas' \citep{Henry2015,Yang2015} with \Lya spectrum taken show in $\sim$ half the galaxies a clear blue peak with always a dominant red size, 
and the $13$ `LARS' galaxies \citep{Ostlin2014,Hayes2014,Rivera-thorsen2014} with \Lya emission suggest have only $\sim $ one clearly detectable blue peak -- which has still more flux on the red side.}. At higher redshifts, lower spectral resolution makes clear distinction in a statistically significant sample harder.
\citet{Erb2014} found in their sample of $36$ LAEs and $122$ LBGs at $z\sim 2-3$ that only $6$ possess a dominant blue side. \citet{Trainor2015} later quantified this number to be $\sim 10\%$ in their extended sample of $350$ LAEs and $65$ LBGs at the same redshift. They also measured in $\sim 45\%$ of the galaxies any blue peak with both fractions being same for their LAE and LBG sample.
Similarly, \citet{Herenz2017} report that $\sim 35\%$ of the $237$ from the MUSE Wide survey show a blue peak, and \citet{2012ApJ...751...29Y} find that approximately half their sample of $91$ \Lya emitting galaxies at $z\sim 3.1$ possess a double peak.

Following this evolution, at higher redshift this ratio might become tilted even more towards the blue. In fact due to a larger gas infall at these early epochs some \Lya radiative transfer studies using cosmological simulations find spectra with all or most of the flux emerging on the blue side \citep[e.g.,][]{Zheng2009,Laursen2009a}. However, since simulations like these fail thus far to reproduce the \Lya spectral properties at lower redshift mentioned above \citep[see comparison by][]{Gronke2018a}, it is yet unclear whether this holds. Also, other observational quantities such as the large \Lya equivalent widths of LAEs at $z\gtrsim 6$ are hard to reconcile with the picture that a majority of the flux (on the blue) is absorbed by the IGM \citep{Matthee2017a}. Nevertheless, the degeneracy between radiative transfer processes on galactic scales, and the IGM do exist (also see discussion in \S~\ref{sec:caveats}). Recently, \citet{Byrohl2020} have suggested a probe to break this degeneracy by the detection of multiple blue peaks which might be feasible in the near future.

In conclusion, it is most likely that our assumption of a symmetric intrinsic spectrum is an over-simplification, and in reality the red peak in $z\gtrsim 6$ galaxies is dominant. This would decrease the number of observable blue peaks in \coda even further. For instance, if the average asymmetries of a \Lya emitters with a blue peak of $\gtrsim 2/3$ of the flux being emitted on the red side (approximately consistent with the lower-$z$ studies mentioned above) holds, we would require $T_{\mathrm{int., blue}}$ close to unity to explain the observations. 

In this study, we compared out findings to three individual galaxies with significant blue flux. Due to a lack of high-resolution \Lya spectra it is too early too say how common such objects are, however, \citet{Songaila2018} state that ``roughly a quarter (two out of eight) of the LAEs have complex profiles with apparent blue wings'' (note, however, that observations carried out using the `Hyper-suprime cam' on the Subaru telescope suggest a lower blue peak fraction at high-$z$, although, no thorough analysis has been performed; \citealp{2014ApJ...785...64S,2019ApJ...883..142H}). While this number is $\sim$ an order of magnitude larger than what we find here, there might be other reasons behind some of the detections.

As mentioned in \citet{Matthee2018}, it is also thinkable that the observations of a double peak at high-$z$ is not a red and a blue peak but instead two red peaks. This could be due to either two intrinsically emitted red peaks, or a wide red peak plus an absorption feature. While double backscattering causes nominally two red peaks\footnote{\new{This is due to the fact that in a `backscattering' event a \Lya photon's frequency is boosted by $\sim 2 v$ where $v$ is the bulk velocity of the scattering medium.}}, the separation of these peaks is too small to yield two distinctive peaks; instead, they are blended causing merely an additional ``bump'' towards the red \citep[as, e.g., in][]{2020MNRAS.492.1778M}. In addition, in such a scenario the peak towards the blue is expected to be stronger.

\begin{figure}
  \centering
  \includegraphics[width=\linewidth]{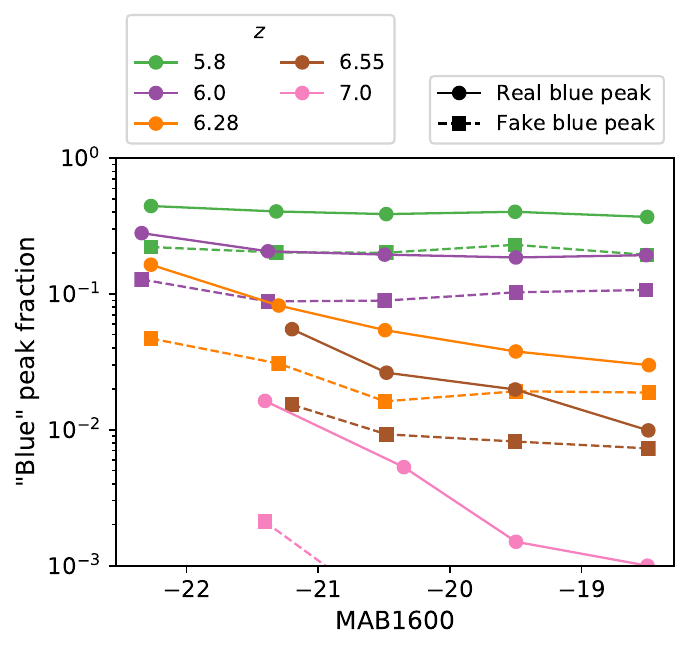}
  \caption{Fraction of blue peaks detectable. The \textit{circles} and \textit{solid} lines show the fraction of `real' blue peaks, i.e., in the range $v\in[-50,\,-400]\kms$, the \textit{squares} connected by \textit{dashed} lines show `fake' blue peaks, i.e., when a intervening absorber could have caused a double peaked detection -- even with two red peaks. See \S~\ref{sec:disc_blueobs} for details.}
  \label{fig:data_forecast}
\end{figure}

However, the second case, i.e., the wide emission plus an absorption, might be more feasible and is in fact the case discussed extensively in \citet{Matthee2018}. While at lower redshift a similar spectrum has not been observed\footnote{Detections of indisputably triple peaked spectra are very rare. The ``Sunburst Arc'' \citep{Rivera-Thorsen2019} and ``Ion2'' \citep{Vanzella2019} are two recent examples from $z\sim 2-3$ \citep[also see][for two more examples at different $z$]{2018MNRAS.476L..15V,2018MNRAS.478.4851I}. Both exhibit a peak at line-center, though, and thus are of different nature than the double peaked detections at $z\gtrsim 6$ but the at least spectrum of ``Ion 2'' could likely be altered by intervening absorbers.} the increasing number density of absorbing systems towards higher redshift \citep{2010ApJ...721.1448S,2012A&A...547L...1N,2015MNRAS.452..217C} might account for this.
In Fig.~\ref{fig:data_forecast} (as squares and dashed lines), we show the likelihood of such an event occurring leading to the observation of a `blue' peak. Specifically, we demand that the red (blue) side is at least $150\kms$ ($100\kms$) wide with a transmission of $T(v) > 0.7$ ($>0.3$) and separated by at least $30\kms$ and a maximum of $300\kms$\footnote{Note that the separation here is not exactly between the peaks but the distance in velocity space between the points where the transmission falls below and raises above the $0.7$ and $0.3$ thresholds, respectively.}. Clearly, given the number of parameters involved in such a scenario, this is merely an example, \new{and we chose the parameters to be conservative, i.e., demanding more realistic (wider) peaks or a larger peak separation will lower the number of `fake' blue peaks.}
However, Fig.~\ref{fig:data_forecast} shows that \new{even with this choice of parameters} such an event occurring is quite unlikely at all redshifts considered, especially towards higher-$z$. Nevertheless, for $z\lesssim 6$ the probability is non-negligible highlighting the importance of systemic redshift measurements which can distinguish between these scenarios \citep[see discussion in][]{Matthee2018}.

Figure~\ref{fig:data_forecast} also shows the probability of observing a `real' blue peak, i.e., at $v<0$ with the same parameters as above. As discussed before, at $z\gtrsim 6.5$ the likelihood is in the percent-level. Noteworthy is the dependence on UV magnitude with a higher probability to see a blue peak in larger halos. After the study of several individual skewers, we attribute this to the fact that a larger surrounding is cut out around these objects due to a larger virial radius; implying whether or not this effect is real is linked to the CGM problematic discussed in \S~\ref{sec:caveats}. Furthermore, we note that in all the cases where a blue peak would be visible, the transmission curve is still a step function as discussed previously but with $v_{\mathrm{cutoff}} < 0$ due to relative motions between the emitting regions and the gas at larger radii. Whether this is realistic, e.g., due to large scale outflows, or infalling galaxies into a potential well, depends highly on \textit{(i)} what the emitting regions are, and \textit{(ii)} how much radiative transfer processes act altering the surface brightness. While we do not address this in detail in this study (see \S~\ref{sec:caveats}), this is an interesting direction for future studies.

\subsection{Comparison to previous work}
\label{sec:disc_previous_work}

Previous work on the impact of the IGM on the \Lya line shape can be grouped into three categories:
\begin{enumerate}
\item (semi-)analytical work which uses a simplified model of the IGM to compute the transmission spectrum. Notable examples of this category are, e.g., \citet{Santos2004} or \citet{Dijkstra2007a} which found the importance of cosmological infall leading to absorption on the red part of the spectrum. They also highlight the impact of the size of the ionized region around \Lya emitters -- which they usually assume to be fully ionized. 
\item work which extends on the first category by not assuming certain `bubble sizes' but using the input of semi-numerical simulations based on an excursion set method such as $21$cmFAST \citep{2011MNRAS.411..955M} in order to map halo masses and redshift to bubble sizes. This approach is very powerful as it is relatively fast, and can, thus, be used to map observations of \Lya observables such as the EW distribution to different reionization histories and eventually constrain the global ionization fraction as, e.g., done in \citet{Mason2017,Mason2019a,Mason2019}. One can combine dark-matter only or hydrodynamic cosmological simulations with ionizing radiative transfer in post-processing  or using the semi-analytical ionization techniques \citep[as recently done in, e.g.,][]{Weinberger2018}. Importantly, in these later, semi-numerical models the ionized regions are set to a constant ionized fraction (usually set by photoionization equilibrium \citealp{Meiksin2009}).
\item There has also been an approach which simulated the EoR by post-processing large, cosmological N-body simulations \citep[e.g.,][]{2007MNRAS.381...75M,2008MNRAS.391...63I,2013MNRAS.428.1366J} or hydrodynamical simulations \citep[e.g.,][]{Laursen2011,2019A&A...627A..84L}. 
 Commonly, ionizing luminosities are assigned to each N-body halo and ray-traced across the density field of the intergalactic gas outside haloes.
  \citet{2008MNRAS.391...63I}, e.g., used this approach to analyze the effect of intergalactic \Lya transmission on LAE observations in a simulation box more than $\sim$100 Mpc on a side, confirming the effect of intergalactic infall surrounding massive galaxies mentioned above.
  They also found that HII regions surrounding
bright LAEs were filled 
with lower-mass halos clustered around the central galaxy
which were also important ionization sources. They note that these `proximity zones' can lead to transmission on the blue side by $z\lesssim 7$.
\item Lastly, there is the approach using a full radiation-hydrodynamics simulation as input, as done in this study \new{\citep[also see,e.g., recent work by][]{2021MNRAS.504.1902G,2021arXiv210510770P}}. Naturally, other studies employed this approach before us \citep[e.g.,][]{Gnedin2004} using different simulations as input.
An advantage of these studies is a more realistic
ionization morphology on small-scales affected by hydrodynamical
backreaction, and better-resolved fluctuations inside the
ionized regions. A disadvantage of these earlier studies is that
they come from much smaller simulation domains, too small to model
the large-scale patchiness of reionization realistically, 
and with resolution too limited to resolve the smaller-mass halos
that can contribute significantly to reionization, even in the neighborhood
of bright LAEs.
\end{enumerate}
  Overall, the findings of these previous studies are fairly consistent (e.g., in pointing out the large sightline-to-sightline variation, or the effect of infall) with differences owing to the specific choice of parameters. Previous work did not, to our knowledge, discuss the effect the IGM transmission on the blue side has on the observable \Lya line properties and, thus, the associated importance of the residual neutral fraction -- and its fluctuations -- of the ionized regions in observing objects like COLA1 -- but instead focused more on the global observables such as the LAE clustering, \Lya luminosity function, and EW distributions, and the effect the intrinsic line shape has on these statistical measures \citep[see, in particular,][who used Gaussian as well as double peaked profiles as intrinsic spectrum]{2013MNRAS.428.1366J}. 

  Clearly, all of the approaches have their own advantages and disadvantages and bring progress in different ways. For instance, recently \citet{2020arXiv200413065M} analyzed the importance of the residual neutral fraction, and its impact on the blue side of the \Lya spectrum, in a simplified model (category \textit{(i)} or \textit{(ii)} above) and demonstrate under which conditions a blue peak is observable.

As we argue in \S~\ref{sec:disc_blueobs}, the occurrence  of blue peaks at high-$z$ is an interesting new way of testing simulations of the EoR against observations.
Current full radiation-hydrodynamics simulations with focus on reionization, include large boxes -- to the sacrifice of spatial resolution -- such as \coda, and smaller ($\lesssim 10$cMpc) boxes focusing on a selected number of halos with higher spatial resolution \citep[e.g.][]{2018MNRAS.479..994R,2019MNRAS.488..419W}. In principle, all these simulations can be tested against \Lya observables, and in particular the occurrence of blue peaks.
We see, however, three main obstacles which we want to caution against: \textit{(i)} although some of the simulations have better resolution, as already discussed above this is thus far still not enough to achieve convergence in \HI (circum-)galactic properties, and hence to resolve structures likely relevant for full \Lya radiative transfer, \textit{(ii)} naturally, in order to compare with observations of galaxies residing in more massive halos such as `Cola1' (i.e., $M_{\mathrm{UV}}\lesssim -21$ at $z \gtrsim 6$), a statistical relevant sample of such halos is required setting a minimum boxsize, and \textit{(iii)} as we found the \Lya transmission (in particular on the blue side) is sensitive to the fiducial neutral fraction which, hence, needs to be captured correctly by simulations in order to use this observable as a probe of the EoR.
This raises a potential issue with the commonly used `reduced speed of light approximation' which is a numerical ingenuity to decrease the computational cost of radiative transfer in simulations but affecting the residual neutral fraction after overlap, as shown in \citet{2019A&A...626A..77O}.

In this wide landscape of theoretical realizations, \coda 's main advantages are its size and the use of the full speed of light.
What is striking, though, is that \coda is too transparent as compared to the \citet{2006AJ....132..117F} measurements (as shown and discussed in \citealp{CoDaII}) but still not transparent enough to yield a large abundance of blue-peaked LAEs. Reproducing both of these aspects has never been done, and seems difficult, as we can judge from this study. It will certainly be an interesting challenge for future numerical simulations of the EoR.

\section{Conclusion}
\label{sec:conclusion}

We analyzed the \Lya transmission properties of the \coda simulation which is a modern cosmological radiation-hydrodynamics simulation. Our findings can be summarized as follows:
\begin{enumerate}
\item the transmission of blue \Lya flux rapidly declines with increasing redshift, due to residual neutral gas inside ionized bubbles which can lead to complete absorption for $n_{\rm HI}\gtrsim 10^{-9}\cm^{-3}$ at $z\gtrsim 7$,
\item there is large sightline variation in blue flux transmission, but no clear $M_{\mathrm{UV}}$ dependence. This is mainly due to kinematic effects, i.e., outflows and / or a relative motion of the emitting galaxy to the surrounding IGM gas. Whether this holds in reality depends strongly on the circumgalactic gas, which we ignore in our analysis, as discussed in \S~\ref{sec:caveats}.
\item the transmission on the red side is greater than the blue side, but also has high sightline variation, in particular at $z \gtrsim 6.5$ when the transmission can vary from zero to unity for a given galaxy.
\item the observed prevalence of blue peaks can provide an additional test for reionization simulations, but better observational statistic are required in order to do so. In \coda, we find for $M_{\mathrm{1600AB}}\sim -21$ galaxies, the opacity of the IGM allows the transmission of blue peaks through $\sim 20\%$ ($\sim 1\%$) of lines-of-sight at $z\sim 6$ ($z\sim 7$).
\end{enumerate}

\section*{Acknowledgments}
\new{The authors thank the referee for constructive feedback that improved the outcome of this study.}
We are grateful to Antoinette Songaila Cowie for sharing the `NEPLA4' spectrum with us.
This research has made use of NASA's Astrophysics Data System, and many open source projects such as \texttt{trident} \citep{Hummels2017}, \texttt{IPython} \citep{PER-GRA:2007}, \texttt{SciPy} \citep{scipy}, \texttt{NumPy} \citep{numpy}, \texttt{matplotlib} \citep{Hunter:2007}, \texttt{pandas} \citep{pandas}, and the \texttt{yt-project} \citep{Turk:2011}.
MG was supported by NASA through
the NASA Hubble
Fellowship grant HST-HF2-51409 awarded by the Space Telescope Science
Institute, which is operated by the Association of Universities for Research in Astronomy, Inc., for NASA, under contract NAS5-26555.
MG acknowledges support from NASA grants HST-GO-15643.017, and HST-AR-15797.001 as well as XSEDE grant TG-AST180036.
CAM acknowledges support by NASA Headquarters through the NASA Hubble Fellowship grant HST-HF2-51413.001-A.
PRS was supported in part by U.S. NSF grant AST-1009799, NASA grant NNX11AE09G,
and supercomputer resources from NSF XSEDE grant TG-AST090005 and the
Texas Advanced Computing Center (TACC) at The University of Texas at Austin.
JM acknowledges a Zwicky Prize Fellowship from ETH Zurich.
GY acknowledges financial support by MICIU/FEDER under project grant
PGC2018-094975-C21.
SEIB acknowledges funding from the European Research Council (ERC) under the European Union’s Horizon 2020 research and innovation programme (grant agreement No. 669253).
ITI was supported by the Science and Technology Facilities Council [grants ST/I000976/1, ST/F002858/1, ST/P000525/1, and ST/T000473/1]; and The Southeast Physics Network (SEPNet).
KA was supported by NRF-2016R1D1A1B04935414 and NRF-2016R1A5A1013277. KA also appreciates APCTP for its hospitality during completion of this work.
PO acknowledges support from the French ANR funded project ORAGE (ANR-14-CE33-0016). ND and DA acknowledge funding from the French ANR for project ANR-12-JS05- 0001 (EMMA). The CoDa II simulation was performed at Oak Ridge National Laboratory/Oak Ridge Leadership Computing Facility on the Titan supercomputer (INCITE 2016 award AST031). Processing was performed on the Eos and Rhea clusters. Resolution study simulations were performed on Piz Daint at the Swiss National Supercomputing Center (PRACE Tier 0 award, project id pr37). The authors would like to acknowledge the High Performance Computing center of the University of Strasbourg for supporting this work by providing scientific support and access to computing resources. Part of the computing resources were funded by the Equipex EquipMeso project (Programme Investissements d’Avenir) and the CPER Alsacalcul/Big Data.

\section*{Data availability}
Data related to this work will be shared on reasonable request to the corresponding author.

\bibliographystyle{mnras}
\bibliography{refs}

\appendix

\section{Changing the fiducial neutral fraction}
\label{sec:fid_neutral_frac}
\begin{figure}
  \centering
  \includegraphics[width=\linewidth]{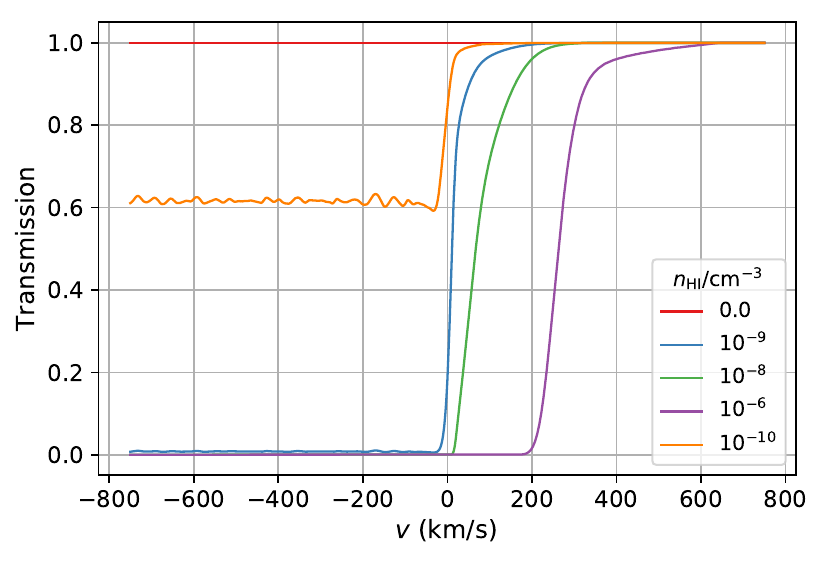}
  \caption{Transmission curves for an homogeneous IGM at $z\sim 7$.}
  \label{fig:fiducial_neutral}
\end{figure}

Figure~\ref{fig:fiducial_neutral} shows transmission curves for homogeneous neutral hydrogen number densities, and no peculiar motion at $z\sim 7$ (but a unchanged temperature). The estimate of $n_{\HI}\sim 2\times 10^{-10}\tau\cm^{-3}$ (cf.  Sec.~\ref{sec:result}) fits this result fairly well.

\bsp	
\label{lastpage}
\end{document}